\documentclass[aps,twocolumn,prresearch,tightenlines,floatfix,superscriptaddress]{revtex4-2}

\usepackage[breaklinks=true]{hyperref}
\usepackage{graphicx}
\usepackage[english]{babel}
\usepackage{amsmath}
\usepackage{amssymb}
\usepackage{times}

\newcommand{\xik}{\xi_{\bf k}}

\begin{document}

\title{Extracting the pairing gap from van Hove singularities in rf spectra of the Fermi Hubbard model}

\author{Chuping Li}
\affiliation{Hefei National Research Center for Physical Sciences at the Microscale and School of Physical Sciences,
University of Science and Technology of China, Hefei, Anhui 230026, China}
\affiliation{Shanghai Research Center for Quantum Science and CAS Center for Excellence in Quantum Information and Quantum Physics,
University of Science and Technology of China, Shanghai 201315, China}
\affiliation{Hefei National Laboratory, University of Science and Technology of China, Hefei 230088, China}

\author{Kaichao Zhang}
\affiliation{Hefei National Research Center for Physical Sciences at the Microscale and School of Physical Sciences,
University of Science and Technology of China, Hefei, Anhui 230026, China}
\affiliation{Shanghai Research Center for Quantum Science and CAS Center for Excellence in Quantum Information and Quantum Physics,
University of Science and Technology of China, Shanghai 201315, China}
\affiliation{Hefei National Laboratory, University of Science and Technology of China, Hefei 230088, China}

\author{Junru Wu}
\affiliation{Hefei National Research Center for Physical Sciences at the Microscale and School of Physical Sciences,
University of Science and Technology of China, Hefei, Anhui 230026, China}
\affiliation{Shanghai Research Center for Quantum Science and CAS Center for Excellence in Quantum Information and Quantum Physics,
University of Science and Technology of China, Shanghai 201315, China}
\affiliation{Hefei National Laboratory, University of Science and Technology of China, Hefei 230088, China}

\author{Yuxuan Wu}
\affiliation{Hefei National Research Center for Physical Sciences at the Microscale and School of Physical Sciences,
University of Science and Technology of China, Hefei, Anhui 230026, China}
\affiliation{Shanghai Research Center for Quantum Science and CAS Center for Excellence in Quantum Information and Quantum Physics,
University of Science and Technology of China, Shanghai 201315, China}
\affiliation{Hefei National Laboratory, University of Science and Technology of China, Hefei 230088, China}

\author{Dingli Yuan}
\affiliation{Hefei National Research Center for Physical Sciences at the Microscale and School of Physical Sciences,
University of Science and Technology of China, Hefei, Anhui 230026, China}
\affiliation{Shanghai Research Center for Quantum Science and CAS Center for Excellence in Quantum Information and Quantum Physics,
University of Science and Technology of China, Shanghai 201315, China}
\affiliation{Hefei National Laboratory, University of Science and Technology of China, Hefei 230088, China}

\author{Pengyi Chen}
\affiliation{Hefei National Research Center for Physical Sciences at the Microscale and School of Physical Sciences,
University of Science and Technology of China, Hefei, Anhui 230026, China}
\affiliation{Shanghai Research Center for Quantum Science and CAS Center for Excellence in Quantum Information and Quantum Physics,
University of Science and Technology of China, Shanghai 201315, China}
\affiliation{Hefei National Laboratory, University of Science and Technology of China, Hefei 230088, China}

\author{Lin Sun}
\email[Corresponding author: ]{lsun22@ustc.edu.cn}
\affiliation{Hefei National Laboratory, University of Science and Technology of China, Hefei 230088, China}
\affiliation{Shanghai Research Center for Quantum Science and CAS Center for Excellence in Quantum Information and Quantum Physics,
University of Science and Technology of China, Shanghai 201315, China}

\author{Qijin Chen}
\email[Corresponding author: ]{qjc@ustc.edu.cn}
\affiliation{Hefei National Research Center for Physical Sciences at the Microscale and School of Physical Sciences,
University of Science and Technology of China, Hefei, Anhui 230026, China}
\affiliation{Shanghai Research Center for Quantum Science and CAS Center for Excellence in Quantum Information and Quantum Physics,
University of Science and Technology of China, Shanghai 201315, China}
\affiliation{Hefei National Laboratory, University of Science and Technology of China, Hefei 230088, China}

\date{\today}

\begin{abstract}
We show that van Hove singularities in rf spectra of the 3D attractive Fermi Hubbard model provide a robust route to extracting the pairing gap. Four types of singularities are classified, and their spectral positions are shown to depend solely on the pairing gap $\Delta$ and chemical potential $\mu$ through simple algebraic relations. Measuring two well-resolved singularities therefore determines both parameters without requiring full spectral fitting. Numerical simulations incorporating phenomenological lifetime and scattering broadenings confirm that these features remain visible in both momentum-integrated and $k_z$-integrated spectra, and become more pronounced at stronger coupling where conventional back-bending methods lose sensitivity. At half filling, particle-hole symmetry fixes $\mu$, reducing the extraction to a single singularity measurement. These results establish vHS analysis as a practical spectroscopic diagnostic for pairing in quantum-simulated 3D Fermi Hubbard systems.
\end{abstract}

\maketitle

\section{Introduction}

Ultracold Fermi gases in optical lattices have garnered significant attention over the past two decades, as their high degree of tunability renders them an ideal platform for quantum simulation and engineering. These systems offer substantial potential for advancing our understanding of long-standing problems in condensed matter physics and exploring novel regimes of quantum matter
\cite{CHEN20051,bloch2008RMP,zwerger2011,bloch2012quantum,hart2015observation,Gross2017}.
The versatility of these platforms arises from the precise control over parameters such as interaction strength, lattice depth, temperature, dimensionality, population imbalance, and lattice geometry.
In particular, the transition from Bardeen-Cooper-Schrieffer (BCS) superfluidity to Bose-Einstein condensation (BEC) of fermion pairs, namely the BCS-BEC crossover, can be realized by tuning the effective interaction strength through a Feshbach resonance in trapped atomic Fermi gases
\cite{jaksch1998cold,chin2010feshbach}. 
This crossover is intimately linked to the pseudogap phenomenon observed in cuprate superconductors~\cite{Timusk1999RPP}, a topic of paramount importance for elucidating the mechanisms of high-temperature superconductivity~\cite{chen1998PRL,RevModPhys.96.025002}.

The physics of ultracold Fermi gases in a three-dimensional optical
lattice (3DOL) can be described by the Fermi Hubbard model.
However, despite extensive theoretical effort~\cite{PhysRevLett.62.1407,PhysRevB.64.075116,PhysRevLett.92.170403,PhysRevA.92.013601,PhysRevLett.98.216402,PhysRevB.69.184501,PhysRevB.78.134503,Micnas1990}, the phase diagram of the 3D attractive FHM at intermediate and strong coupling, especially at high fillings, remains largely undetermined. In the weak-coupling limit, BCS mean-field theory, along with the Gor'kov correction \cite{gor1961}, provides a reliable description. In the BEC regime at low densities, the physics reduces to a dilute gas of tightly bound pairs whose few-body properties are well understood. Away from these limiting regimes, however, in the intermediate coupling and density range where pairing fluctuations are the most prominent, diagrammatic methods capture qualitative trends but differ quantitatively in their predictions for key observables such as the pairing gap $\Delta$. For instance, the non-self-consistent $G_0G_0$ approximation \cite{nozieres1985bose}, widely used for its computational simplicity, the $G_0G$ scheme (derived using an equation of motion approach \cite{Kadanoff1961,Chen2000}, employed in the present work), and the conserving $GG$ theory \cite{Tchernyshyov1997,haussmann1994,Haussmann2009} each yield appreciably different values of $\Delta$ in the pseudogap regime where the effects of pairing fluctuations are strong. Often regarded as a reliable numerical method, quantum Monte Carlo (QMC), which is free of the fermion sign problem in the attractive case, is limited by the analytic continuation needed to obtain real-frequency spectral information, which obscures fine spectral features at strong coupling. In addition, it suffers from numerical instability at strong coupling and becomes computationally prohibitive at large lattice sizes and low temperatures. Consequently, quantum simulation experiments that realize the 3D attractive FHM in controllable cold-atom platforms~\cite{schneider2012fermionic,mitra2018quantum,hackermuller2010anomalous,PhysRevLett.124.010403,brown2020angle} are indispensable for establishing the phase diagram, and recent progress, including measurements of the doublon fraction~\cite{Zhu2026}, underscores the need for complementary diagnostics that can directly extract pairing parameters from spectroscopic measurements. This work develops such a diagnostic.

In particular, radio-frequency (rf) spectroscopy serves as an important probe for
investigating superfluid pairing physics and has been widely used in
experiments on ultracold Fermi
gases~\cite{Torma_2016,Vale2021}. Initially utilized to probe
mean-field shifts in strongly interacting systems, the technique
evolved following the proposal that the pairing gap could be extracted
spectroscopically~\cite{PhysRevLett.85.487}. Subsequent experiments
have provided compelling evidence of fermionic pairing through
distinct spectral
signatures~\cite{Chin2004,schunck2008determination,PhysRevLett.101.140403}. Nevertheless,
the peak position in momentum-integrated rf spectra can hardly be used
to extract the pairing gap, since the fermion energy shift induced by
pairing is momentum-dependent and not proportional to the gap~\cite{Chen_2009}.

Momentum-resolved rf spectroscopy, analogous to angle-resolved photoemission spectroscopy (ARPES), overcomes this limitation by directly measuring the spectral function $A(\mathbf{k}, \omega)$~\cite{Stewart2008,Gaebler2010,PhysRevLett.114.075301}. In homogeneous continuum gases, the pairing gap can then be inferred from the characteristic back-bending of the dispersion, a technique that has recently confirmed the presence of a pseudogap in strongly interacting Fermi gases~\cite{li2024observation}.

In optical lattices, however, the anisotropic dispersion makes the direct reconstruction of full 3D momentum-resolved rf spectra experimentally demanding. Such measurements have been successfully demonstrated in 2D lattice systems~\cite{brown2020angle}, but extending them to three dimensions remains beyond reach at present. Theoretical studies of rf spectroscopy in the attractive Hubbard model have also been limited~\cite{Datta2014}. Consequently, it is desirable to identify spectral markers that survive partial momentum integration and whose positions are directly linked to the underlying pairing parameters.

In this work, we investigate the effects of van Hove singularities (vHSs) in both momentum-integrated and momentum-resolved rf spectra for two-component ultracold Fermi gases in a 3DOL with nearest-neighbor hopping, described by an attractive Fermi Hubbard model. The vHS features originate from the underlying lattice band structure and are subsequently modified by fermion pairing. We show that, within the present approximation, the spectral positions of the vHS features are controlled by the pairing gap $\Delta$ and the effective chemical potential $\mu$. In principle, measuring two well-resolved vHS positions therefore allows one to infer both $\Delta$ and $\mu$ from the corresponding algebraic relations. This strategy is complementary to the conventional analyses based on rf spectral peaks and back-bending of the quasiparticle dispersion, because the vHS positions are fixed by lattice critical points, they remain robust and identifiable even when the momentum is only partially resolved in the rf spectra.
Employing the $G_{0}G$ $T$-matrix approximation within the framework of pair-fluctuation theory~\cite{PhysRevB.59.7083}, we solve the 3D attractive FHM at finite temperature $T$ and fermion density $n$ for different interaction strengths $U$.
We introduce two phenomenological broadening parameters into the self-energy to simulate momentum-integrated and momentum-resolved rf spectra under realistic experimental conditions~\cite{Chen2001,Chen2009}.
Because momentum-resolved rf measurements typically involve integration over one momentum component (line-of-sight integration), we also calculate the $k_z$-integrated momentum-resolved spectrum and identify the four types of vHSs in these rf spectra, each appearing at distinct rf detuning positions. 
For illustrative purposes, we keep the main text focused on the representative cases $n=0.1$ and $n=1$. Additional results for intermediate fillings ($n=0.2,0.4,0.6,0.8$) and interaction strengths are shown in the Supplemental Material~\cite{supplemental_material}. The detailed self-consistent equations are also provided there. 

\section{\label{sec:1}Theoretical Formalism}
\subsection{Model and overview of the pairing fluctuation theory}
The attractive FHM in momentum space is described by
\begin{eqnarray}
	H\! - \!\mu \hat{N}
	= \sum_{\mathbf{k}\sigma} \xik \hat{c}_{\mathbf{k}\sigma}^{\dag} \hat{c}_{\mathbf{k}\sigma}
	+ U\!\sum_{\mathbf{k}\mathbf{k}^{'}\mathbf{q}} 
	\hat{c}_{\mathbf{k}_+\uparrow}^{\dag} \hat{c}_{-\mathbf{k}_-\downarrow}^{\dag} \hat{c}_{-\mathbf{k}^{'}_-\downarrow} \hat{c}_{\mathbf{k}^{'}_+\uparrow},\nonumber
\end{eqnarray}
where $\xi_{\mathbf{k}} = \epsilon_{\mathbf{k}} - \mu$,
$\mathbf{k}_\pm\equiv\mathbf{k}\pm\mathbf{q}/2$, and $\mu$ is the
effective fermion chemical potential into which the mean-field Hartree
shift has already been absorbed. Here $U<0$ is the attractive
interaction strength, and $\epsilon_{\bf k}=2t(3-\cos{k_{x}}-\cos{k_{y}}-\cos{k_{z}})$ is the
tight-binding dispersion, where $t$ is the nearest-neighbor hopping
amplitude and the lattice constant is set to unity.
Throughout this work, we take the half bandwidth $6t$ as the energy unit. 

We solve the attractive FHM using the $G_0 G$ scheme of the $T$-matrix approximation of pair-fluctuation theory, which self-consistently includes pairing fluctuations beyond mean-field BCS theory, as derived from the earlier work of Kadanoff and Martin~\cite{Kadanoff1961}. The full formalism is detailed in Refs.~\cite{PhysRevB.59.7083,Chen_2009} and in the Supplemental Material~\cite{supplemental_material}; here we only outline the essential physical structure.

The fermion self-energy $\Sigma(K) = \Sigma_{\rm sc}(K) + \Sigma_{\rm pg}(K)$ contains in general contributions from both zero-momentum pairing (relevant below $T_{\rm c}$, giving rise to the superfluid condensate contribution $\Sigma_{\rm sc}$) and finite-momentum pairing. The latter, referred to as the pseudogap self-energy, is expressed through the $T$-matrix $t_{\rm pg}(Q) = U/[1+U\chi(Q)]$, where $\chi(Q)=\sum_K G_0(Q-K)G(K)$ is the pair susceptibility. Near $T_{\rm c}$, the dominant contribution to $\Sigma_{\rm pg}$ comes from the vicinity of $Q= 0$, allowing the pseudogap to be captured by a single energy scale $\Delta_{\rm pg}^2 = -\sum_{Q\neq 0} t_{\rm pg}(Q)$.
The total pairing gap $\Delta = \sqrt{\Delta_{\rm sc}^2 + \Delta_{\rm pg}^2}$ enters a BCS-like gap equation and particle-number equation, which are solved self-consistently for $\mu$, $\Delta_{\rm sc}$, and $\Delta_{\rm pg}$ at given $U$, $n$, and $T$~\cite{supplemental_material}. The resulting Green's function takes the standard BCS form
\begin{equation}
    \label{eq:Greenfunction}
    G(K) = \frac{u_k^2}{i\omega_l -E_{\bf k}} + \frac{v_k^2}{i\omega_l +E_{\bf k}} \,,
\end{equation}
with coherence factors $u_k^2, v_k^2 = (1\pm \xi_{\mathbf{k}}/E_{\mathbf{k}})/2$ and $E_{\mathbf{k}}=\sqrt{\xi_{\mathbf{k}}^{2}+\Delta^{2}}$.

\subsection{Spectral function and van Hove singularities}

The spectral function $A({\bf k},\omega)$ follows directly from Eq.~(\ref{eq:Greenfunction}) as
\begin{eqnarray}
A({\bf k},\omega)&=&-2\,{\rm{Im}}\,G^{\rm R}({\bf k},\omega)\nonumber\\
&=&2\pi\left[u_{\bf k}^{2}\delta(\omega-E_{\bf k})+v_{\bf k}^{2}\delta(\omega+E_{\bf k})\right]\,,
\label{eq:BCSAkw}
\end{eqnarray}
exhibiting two quasiparticle branches with weights given by the coherence factors.

\begin{figure}
	\centering
	\includegraphics[clip,width=3.4in] {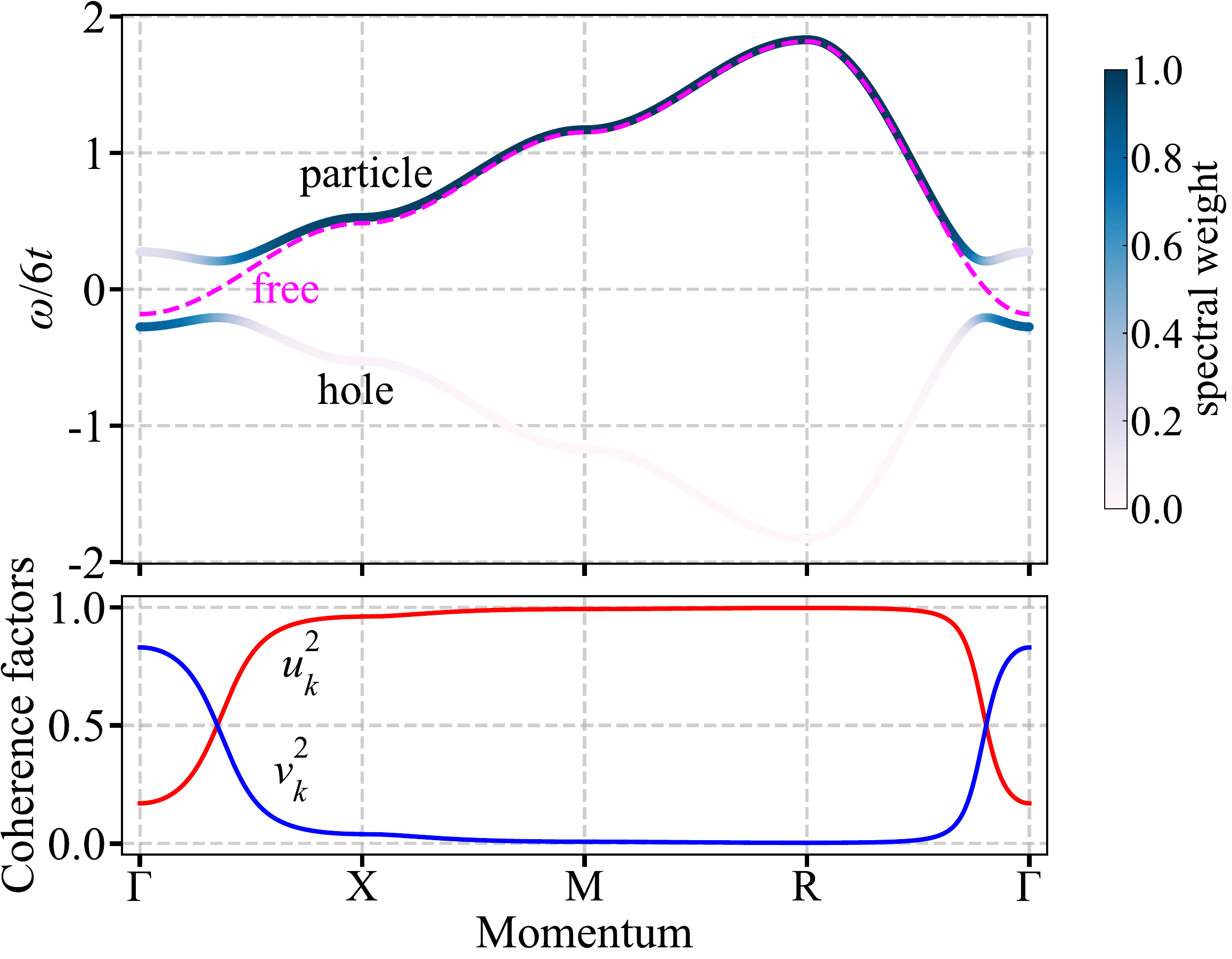}
	\caption{Representative spectral function of the 3D attractive FHM at $U_\text{c}$ and $T_{\rm c}$ for $n=0.1$. The results are obtained from self-consistent solutions yielding $\Delta=0.207$ and $\mu=0.182$ (in units of $6t$). The upper and lower blue curves denote the particle and hole quasiparticle branches, respectively, with the two branches color-coded by their spectral weight. The dispersions are plotted along the high-symmetry lines of the first Brillouin zone, with $\Gamma=(0,0,0)$, $\text{X}=(\pi,0,0)$, $\text{M}=(\pi,\pi,0)$, and $\text{R}=(\pi,\pi,\pi)$. The magenta curve represents the noninteracting band dispersion. The lower panel shows the coherence factors $u^{2}_{\bf k}$ and $v^{2}_{\bf k}$, corresponding to the spectral weights of the two quasiparticle branches. }
	\label{fig:Akwinitial}
\end{figure}

Figure~\ref{fig:Akwinitial} illustrates a typical spectral function for a Fermi gas at critical coupling $U_\text{c} \approx -7.9136t$ in a 3DOL, corresponding to the threshold for forming a two-body bound state with zero binding energy in the dilute limit. The upper and lower branches correspond to the ``particle'' and ``hole'' quasiparticles, respectively.
As depicted in the lower panel, the coherence factors $u^{2}_{\mathbf{k}}$ and $v^{2}_{\mathbf{k}}$ dictate the spectral weights of these quasiparticles at energies $\pm E_{\bf k}$. The locus $\xi_{\mathbf{k}}=0$ for free fermions defines the Fermi surface. 
A comparison between the quasiparticle dispersion and the noninteracting band dispersion reveals a distinct pairing gap opening at the Fermi surface, whose crossing with the quasiparticle dispersion is located midway along the $\Gamma-\text{X}$ and $\Gamma-\text{R}$ paths.

A van Hove singularity, associated with a non-analytic feature in the density of states (DOS) $N(\omega)=\sum_{\mathbf{k}} A({\bf k},\omega)$~\cite{czycholl2023solid}, arises from the vanishing gradient of the lattice dispersion, $\nabla_{\bf k} \xi_{\bf k}=0$. For a standard 3D cubic lattice, the Hessian matrix $\mathcal{H}(\xi_{\mathbf{k}})$ evaluated at the critical points is diagonal:
\[ 
  {\mathcal H}(\xi_{\mathbf{k}})\propto \begin{bmatrix}
    \cos{k_{x}} & 0 & 0\\
    0 & \cos{k_{y}} & 0\\
    0 & 0 & \cos{k_{z}}
  \end{bmatrix}.
\]
When $k_{x,y,z}$ take values $0$ or $\pi$, four distinct types of vHSs arise, including a local minimum ($\mathrm{M_0}$, all cosines positive), a local maximum ($\mathrm{M_3}$, all cosines negative), and two types of saddle points ($\mathrm{M_1}$ and $\mathrm{M_2}$, with mixed signs). Along the dispersion shown in Fig.~\ref{fig:Akwinitial}, the $\Gamma$, $\text{X}$, $\text{M}$, and $\text{R}$ points correspond to these four vHS types, respectively, i.e., $(\Gamma,\text{X},\text{M},\text{R}) \leftrightarrow (\mathrm{M_0},\mathrm{M_1},\mathrm{M_2},\mathrm{M_3})$.

\subsection{rf response and vHS kinematic relations}

Within linear response theory, the lowest-order rf current for a transition from an initial interacting state to a final noninteracting state is~\cite{Chen_2009}
\begin{equation}
    \label{eq:RFcurrent}
    I(\nu)=\frac{1}{2\pi}\sum_{\mathbf{k}} A({\bf k},\omega)f(\omega)\Big|_{\omega=\xi_{\mathbf{k}}-\nu},
\end{equation}
where $\nu$ is the rf detuning. The energy-conservation constraint $\omega=\xi_{\mathbf{k}}-\nu$ reveals a structural resemblance between the rf response and the DOS, suggesting that vHS signatures will also manifest in the rf spectroscopy of lattice Fermi gases. 

At low temperatures, the particle branch is exponentially suppressed by the Fermi function. Focusing on the hole branch and substituting into Eq.~(\ref{eq:RFcurrent}) gives
\begin{equation}
    \label{eq:RFcurrent2}
    I_{h}(\nu)=\int\!\! \frac{\mathrm{d}^3k}{(2\pi)^3} v^{2}_{\bf k} f(\xi_{\mathbf{k}}-\nu)\delta(\xi_{\mathbf{k}}-\nu+E_{\bf k}) \,.
\end{equation}
By imposing the resonance condition $\xi_{\mathbf{k}}-\nu+E_{\bf k}=0$ at the four vHS momenta, the characteristic rf detunings of the vHS are obtained as
\begin{subequations}
\label{eq:fourvHS}
\begin{eqnarray}
{\rm M_0}:& \nu_{0}=&-\mu+\sqrt{\mu^{2}+\Delta^{2}} \,,\\
{\rm M_1}:& \nu_{1}=&\frac{2}{3}-\mu+\sqrt{(\frac{2}{3}-\mu)^{2}+\Delta^{2}} \,,\\
{\rm M_2}:& \nu_{2}=&\frac{4}{3}-\mu+\sqrt{(\frac{4}{3}-\mu)^{2}+\Delta^{2}} \,,\\
{\rm M_3}:& \nu_{3}=&2-\mu+\sqrt{(2-\mu)^{2}+\Delta^{2}}
\,.\label{eq:1234}
\end{eqnarray}
\end{subequations}
The band energies at the vHS points are $\epsilon_i=2i/3$ ($i=0,1,2,3$), i.e., $\epsilon_0=0$, $\epsilon_1=2/3$, $\epsilon_2=4/3$, $\epsilon_3=2$. Equivalently, the four detunings can be combined into the unified expression
\begin{equation}
    \label{eq:unifiedvHS}
    \nu_i = \epsilon_i - \mu + \sqrt{(\epsilon_i-\mu)^{2}+\Delta^{2}}, \qquad i=0,1,2,3,
\end{equation}
with all energies in units of $6t$.
Equations~(\ref{eq:fourvHS}) are the central results of this analysis within the BCS-like quasiparticle description, showing that the rf detuning of each vHS is determined by $\mu$ and $\Delta$ once additional self-energy shifts are neglected or absorbed into these parameters. Since there are only two unknowns, any two vHS positions that are both spectrally resolved and carry sufficient weight can, in principle, be used to solve for $\Delta$ and $\mu$ algebraically. The practical choice of the two markers is therefore constrained by spectral visibility and separation from neighboring structures. In the special case of half filling ($n=1$), the particle-hole symmetry pins $\mu=1$, so that a single measured vHS uniquely determines $\Delta$.

Experimentally, one can achieve momentum resolution in the $xy$-plane
by shining the imaging laser in the $k_z$ direction. This leads to
$k_z$-integrated spectral response, given by integrating
Eq.~(\ref{eq:RFcurrent2}) over $k_z$:
\begin{eqnarray}
    \label{eq:z-integralRF}
    I_{h}(k_{x},k_{y},\nu)&=&\frac{1}{\pi^3}\!\int^{\pi}_{0}\!\mathrm{d}k_{z}\, v^{2}_{\bf k} f(\xi_{\mathbf{k}}-\nu)\delta(\xi_{\mathbf{k}}-\nu+E_{\bf k}) \nonumber\\
    &=&\frac{\Delta^{2}}{\nu^{2}}\left[1-f\left(\frac{\nu^{2}+\Delta^{2}}{2\nu}\right)\right]\frac{3}{|\sin{k_{z_0}}|},
\end{eqnarray}
where $k_{z_0}(k_x,k_y,\nu)$, defined as the root of $\xi_{\mathbf{k}}-\nu+E_{\mathbf{k}}=0$ for given $k_x$, $k_y$ and $\nu$, signals a VHS when $k_{z_0}=0$ or $\pi$, 
and mirror symmetry has been used to fold $\pm k_x$ and $\pm k_y$ so that $(k_x,k_y)$ is restricted to the first quadrant of the first Brillouin zone (BZ). In this projected representation, the vHS effects are most pronounced at the center and the BZ boundaries.

\begin{figure}
	\centerline{\includegraphics[clip,width=3.3in] {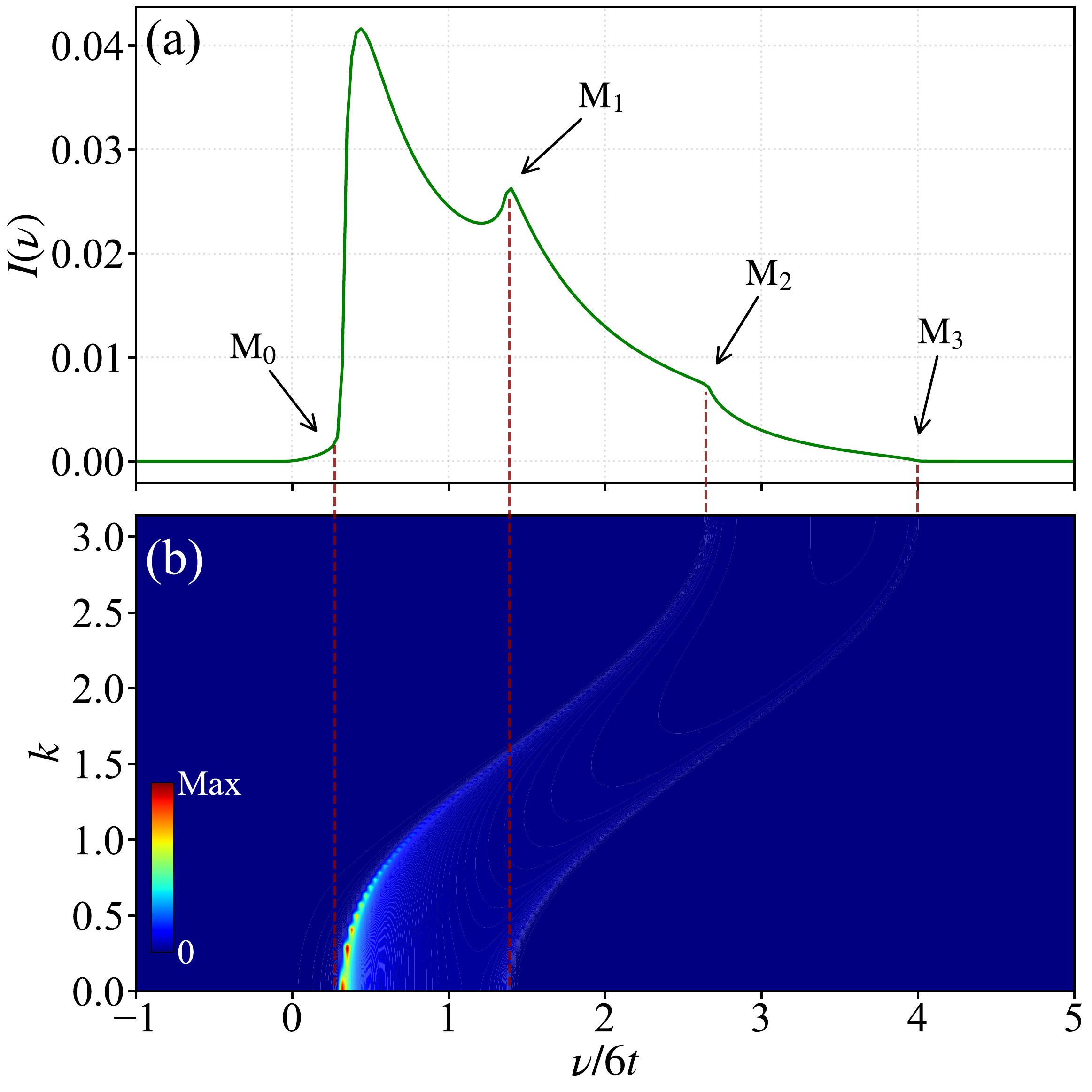}}
	\caption{(a) rf current $I(\nu)$, i.e., the momentum-integrated rf spectrum, at $T_{\rm c}$ for $U/6t=-1.8$ and $n=0.1$. The four types of vHSs are labeled on the curve. (b) The $z$-integrated, partially momentum-resolved rf spectrum $I(k,\nu)$ plotted along the projected face-diagonal direction $\mathbf{k}_{\parallel}=(k,k)$. The red dashed lines align the corresponding vHS features in (a) and (b). }
	\label{fig:RFcurrentandAkxy_w}
\end{figure}

Shown in Fig.~\ref{fig:RFcurrentandAkxy_w} are the energy positions of
the four types of vHSs in the momentum-integrated rf current $I(\nu)$,
and their counterparts in the $k_z$-integrated momentum-resolved
response $I(k,\nu)$ along the face-diagonal direction
$\mathbf{k}_{\parallel}=(k,k)$.  In the momentum-integrated spectra in
Fig.~\ref{fig:RFcurrentandAkxy_w}(a), the $\rm M_1$ vHS is the most
prominent and should be experimentally accessible.  Compared to the
fully integrated signal, the vHS effects in the partially
momentum-resolved spectrum in Fig.~\ref{fig:RFcurrentandAkxy_w}(b) are
significantly more pronounced, because reducing the integration
dimensionality better preserves the non-analytic nature of the
dispersion. Correspondingly, extracting the energy distribution curves
(EDCs) from such partially momentum-resolved rf data is expected to yield
unambiguous experimental signatures of these vHS features.

\subsection{Phenomenological spectral broadening}
In our initial theoretical framework, the spectral function exhibits infinitely sharp quasiparticle peaks. To simulate the finite lifetime of thermally excited pairs and single-particle scattering, we generalize the (retarded) self-energy to~\cite{MALY1999113,PhysRevB.57.R11093,Chen2001}
\begin{equation}
    \label{eq:general_pseudogap}
    \Sigma^{\rm R}_{\rm pg}({\bf k},\omega)=\frac{\Delta^{2}_{\rm pg}}{\omega+\xi_{\mathbf{k}}+\mathrm{i}\Gamma_{0}}-\mathrm{i}\Gamma_{1},
\end{equation}
where $\Gamma_0$ describes the broadening of quasiparticle peaks associated with the finite pair lifetime, and $\Gamma_1$ accounts for the incoherent background from single-particle scattering. The full spectral function follows from Dyson's equation:
\begin{equation}
    \label{eq:generalized_Akw}
    A({\bf k},\omega)\!=\!\frac{-2\,{\rm{Im}}\Sigma^{\rm R}_{\rm pg}}{[\omega\!-\xi_{\mathbf{k}}-\!\Sigma^{\rm R}_{\rm sc}\!-\!{\rm Re}\Sigma^{\rm R}_{\rm pg}]^2\!+\![{\rm{Im}}\Sigma^{\rm R}_{\rm pg}]^2}.
\end{equation}
In the results below, we treat $\Gamma_0$ and $\Gamma_1$ as adjustable phenomenological parameters to systematically assess the visibility and resolvability of the vHS features under  realistic experimental conditions; both parameters become negligible at low temperatures due to the formation of stable pairing~\cite{long_article}. The broadened BCS form of the spectral function underlying Eq.~(\ref{eq:generalized_Akw}) is well supported experimentally, as recent rf measurements in homogeneous unitary Fermi gases~\cite{li2024observation}, together with earlier rf studies~\cite{JinStrinati_nphys}, have confirmed that the spectral function retains a BCS-like two-branch structure even in the strongly interacting regime. We note that the vHS detunings in Eqs.~(\ref{eq:fourvHS}) follow from the resonance condition $\xi_{\mathbf{k}}-\nu+E_{\bf k}=0$ and therefore define the underlying kinematic locations of the vHS features. The phenomenological broadening mainly smears these singular structures and reduces their contrast, and this effect diminishes as the temperature decreases. 
As verified numerically in Sec.~III, the spectral features remain centered close to the kinematic detunings over the broadening range considered here.

\section{\label{sec:2}Numerical Results and Discussion}

\subsection{Spectral function}
\begin{figure}
	\centerline{\includegraphics[clip,width=3.4in] {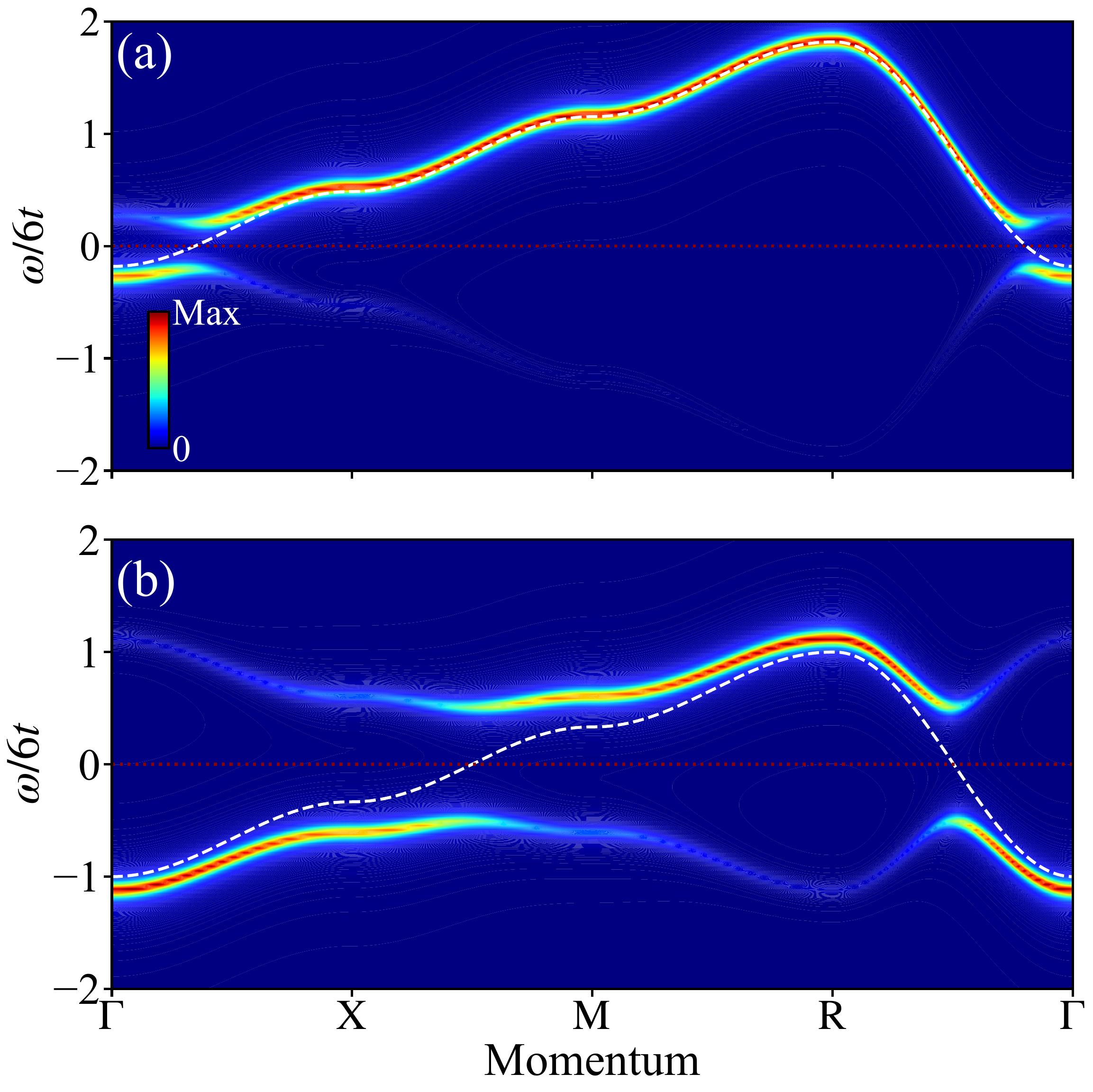}}
	\caption{Spectral function $A({\bf k},\omega)$ of the 3D attractive FHM at $T_{\rm c}$ and $U_{\rm c}$ for (a) $n=0.1$ and (b) $n=1$, with $\Gamma_{0}=\Gamma_{1}=0.05$. The white dashed line denotes the noninteracting band dispersion. }
	\label{fig:AkwGaussconv}
\end{figure}
Based on the self-consistent solutions, we evaluate the spectral function using Eq.~(\ref{eq:generalized_Akw}) with $\Gamma_{0}=\Gamma_{1}=0.05$. Figure~\ref{fig:AkwGaussconv} illustrates the qualitative evolution of $A({\bf k},\omega)$ for different densities (a) $n=0.1$ and (b) $n=1$ along the high-symmetry momentum directions at $U_\text{c}$ and $T_{\rm c}$. Results of more densities can be found in the Supplemental Material~\cite{supplemental_material}. A pronounced gap opens at the Fermi surface for both densities. 
Along the $\Gamma-\text{R}$ direction (the body diagonal), a characteristic back-bending feature occurs at the Fermi wave vector ${\bf k}_{\mu}$ determined by $\xi_{{\bf k}_{\mu}}=0$. At low filling, the spectral function along $\Gamma-\text{R}$ resembles that of the 3D continuum limit, where $k_{\mu}\simeq\sqrt{\mu}$. As $n$ increases, the gap widens and the spectral weight of the lower hole branch is enhanced, with lattice effects becoming increasingly apparent. 

At half filling ($n=1$), particle-hole symmetry fixes the chemical potential at the band center ($\mu=1$). The corresponding Fermi-surface crossing along $\Gamma-\text{R}$ is located at ${\bf k}_{\mu}=(\pi/2,\pi/2,\pi/2)$, i.e., exactly at the midpoint of the $\Gamma-\text{R}$ line, as shown in Fig.~\ref{fig:AkwGaussconv}(b). This special kinematic point provides a useful experimental diagnostic to identify half filling; combined with the vHS protocol proposed below, it enables $\Delta$ to be extracted from a single vHS measurement.

\subsection{rf current and sensitivity to broadening}

The fully momentum-integrated rf current $I(\nu)$ is the simplest experimental observable. We compute $I(\nu)$ for various $(U,n)$ and examine the sensitivity of the vHS features to the broadening parameters $\Gamma_0$ and $\Gamma_1$.

\begin{figure}
	\centerline{\includegraphics[clip,width=3.4in] {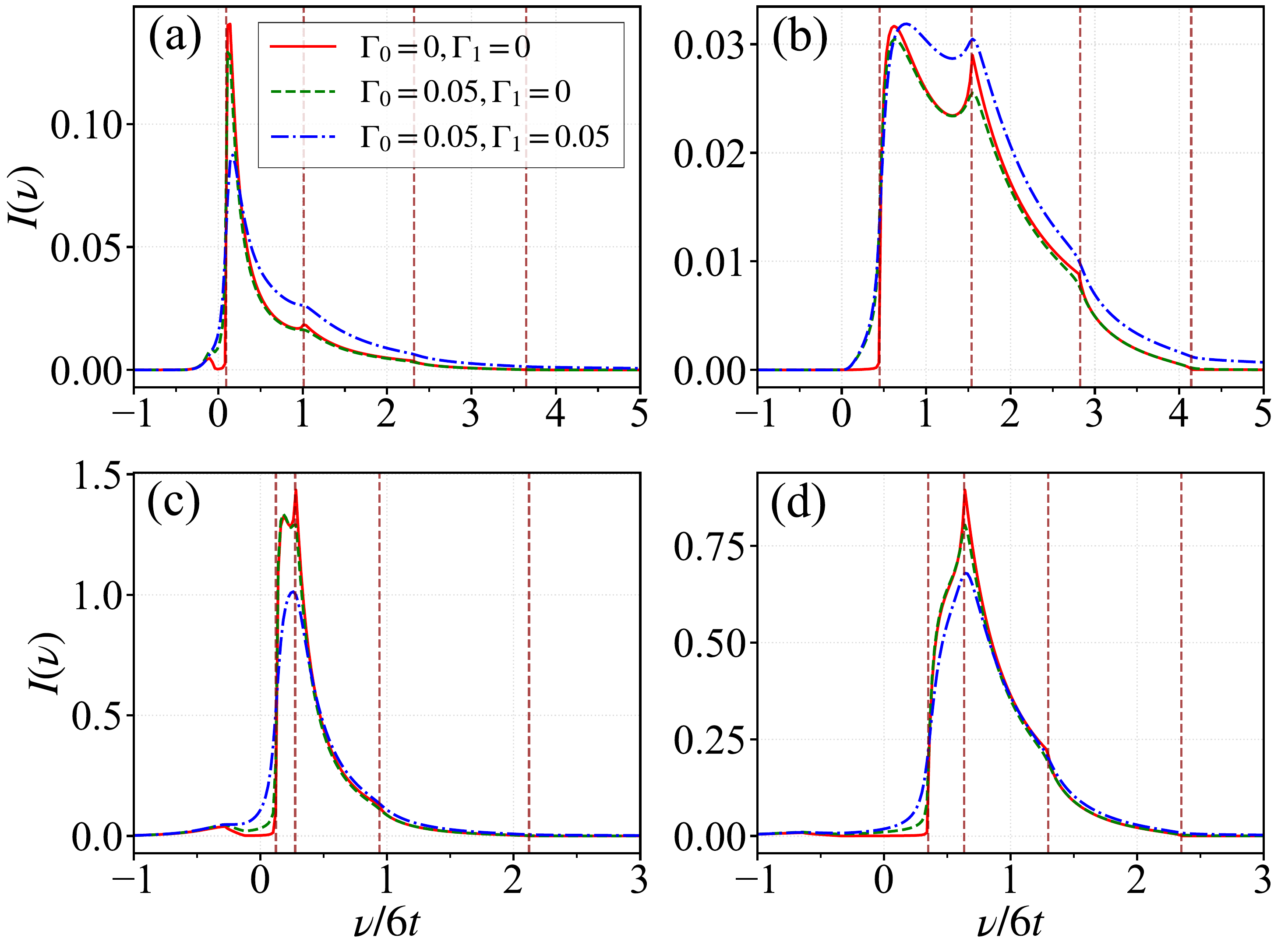}}
	\caption{rf current $I(\nu)$ as a function of detuning $\nu$ for (a) $n=0.1$, $U=U_{\rm c}$, (b) $n=0.1$, $U/6t=-2$, (c) $n=1$, $U=U_{\rm c}$, and (d) $n=1$, $U/6t=-2$, for various $\Gamma_0$ and $\Gamma_1$. The vertical lines indicate the rf detunings corresponding to the $\mathrm{M}_{0}$--$\mathrm{M}_{3}$ vHSs.}
	\label{fig:RFcurrent}
\end{figure}

As shown in Fig.~\ref{fig:RFcurrent}, for a fixed filling, as the coupling increases from weak to strong, the pairing gap grows and the rf spectral weight shifts and spreads toward larger detuning, leading to a broader spectrum consistent with the observations in continuum gases~\cite{schunck2008determination}. The shift simultaneously suppresses the peak near the $\rm M_0$ detuning at the left edge, making the $\rm M_{1}$ vHS feature relatively more pronounced, as shown in Fig.~\ref{fig:RFcurrent}(b). As the density (and chemical potential) increases toward half filling, the left-edge peak and the $\rm M_1$ peak approach each other, and the spectral weight of the former becomes so suppressed that the $\rm M_1$ peak becomes dominant. Upon broadening, the two features merge into a single dominant peak, as seen in Fig.~\ref{fig:RFcurrent}(c) and Fig.~\ref{fig:RFcurrent}(d). Compared with the low-filling case, the spectral peak at high filling is then associated primarily with the $\rm M_1$ vHS, reflecting the proximity of the Fermi level to the vHS band extrema.

The broadenings primarily reduce the contrast of the vHS features and merge nearby structures, while their spectral weight remains concentrated near the kinematic detunings predicted by Eqs.~(\ref{eq:fourvHS}). The apparent peak maxima may change slightly when features overlap, but the underlying vHS markers are only weakly sensitive to $\Gamma_0$ and $\Gamma_1$ within the parameter range at low temperature. This supports the use of Eqs.~(\ref{eq:fourvHS}) as a stable guide for extracting $\Delta$ and $\mu$ from well-resolved spectral features. Note that the $\rm M_0$ detuning, corresponding to the $\Gamma$ point, provides a particularly strong marker in the large-gap regime, where it moves well away from zero detuning.

\subsection{$k_z$-integrated momentum-resolved rf spectroscopy}

The observable more directly accessible in experiments is the $k_z$-integrated, $(k_x,k_y)$-resolved spectrum $I({\bf k}_{\parallel},\nu)$, which retains the momentum resolution in the $xy$ plane. We focus on the face-diagonal direction ${\bf k_{\parallel}}=(k,k)$ and  axis-along direction ${\bf k_{\parallel}}=(0,k)$ with $\Gamma_0=\Gamma_1=0.05$. In contrast to the fully integrated spectrum, where the $\rm M_1$ feature dominates, the partial momentum resolution brings the $\mathrm{M}_2$ and $\mathrm{M}_3$ features to the foreground along the face-diagonal direction. As discussed below, however, the $\mathbf{k}_{\parallel}=(0,\pi)$ points turns out to be more favorable experimentally, since the $\mathrm{M}_1$ and $\mathrm{M}_2$ signals there are considerably stronger.

\begin{figure}
	\centerline{\includegraphics[clip,width=3.4in] {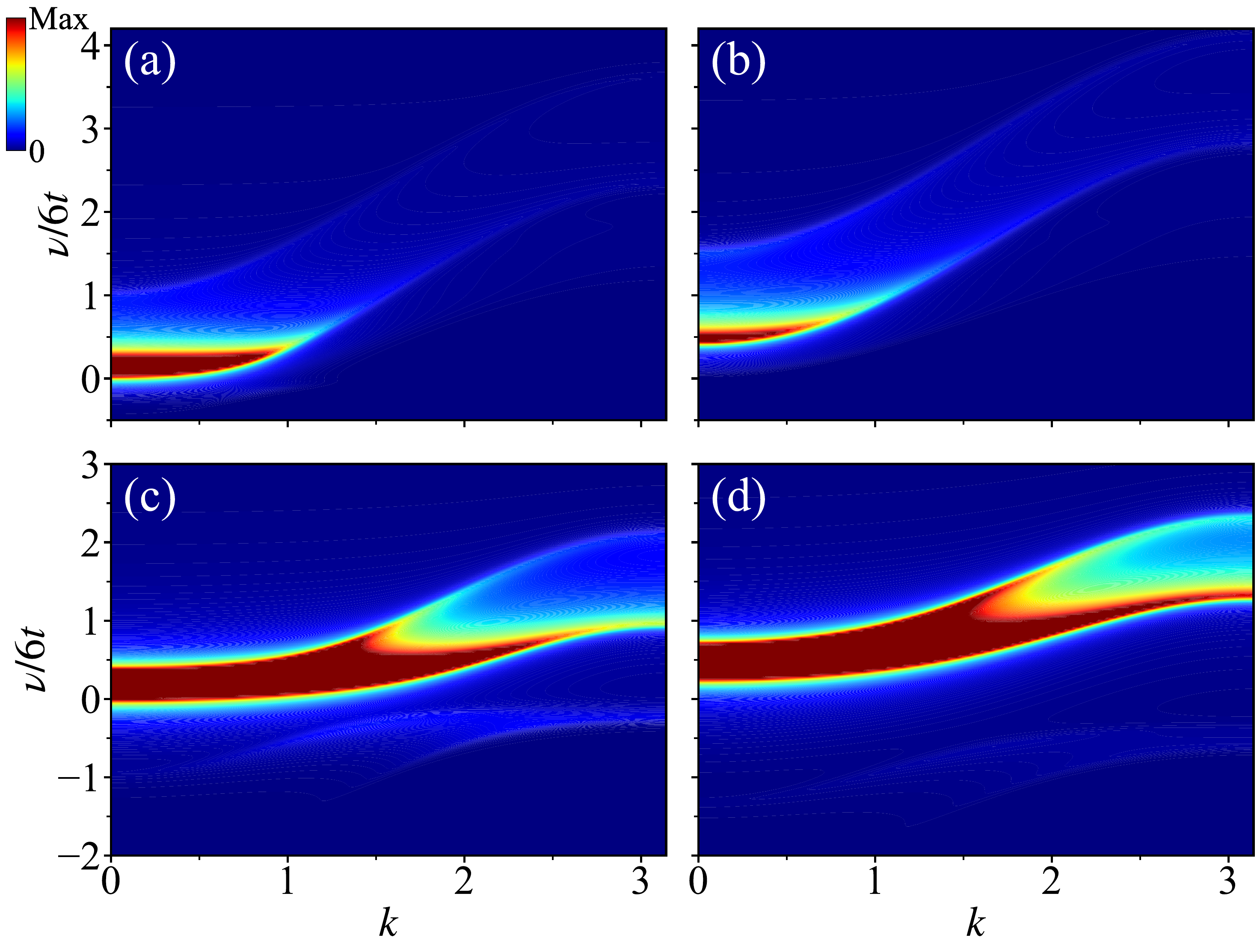}}
	\caption{The $k_z$-integrated, $(k_x,k_y)$-resolved rf spectra $I(k,\nu)$ at $T_{\rm c}$ along the face-diagonal direction ${\bf k}_{\parallel}=(k,k)$ for (a) $n=0.1$, $U=U_{\rm c}$, (b) $n=0.1$, $U/6t=-2$, (c) $n=1$, $U=U_{\rm c}$, and (d) $n=1$, $U/6t=-2$. The same color saturation value is applied across all panels to enhance the visibility of weak signals. }
	\label{fig:z-integrated_MRRF}
\end{figure}

Comparing Fig.~\ref{fig:z-integrated_MRRF}(a) and Fig.~\ref{fig:z-integrated_MRRF}(b), increasing $|U|$ shifts the dominant spectral weight toward larger rf detunings, consistent with a larger pairing gap. When the filling increases to $n=1$ [Fig.~\ref{fig:z-integrated_MRRF}(c) and Fig.~\ref{fig:z-integrated_MRRF}(d)], the spectral weight also shifts toward larger detunings as the pairing gap grows at these interaction strengths, while the spacing between the vHS features decreases. As shown in Fig.~\ref{fig:AkwGaussconv}, this behavior reflects the reduced curvature of the dispersion near the relevant band extrema; as the Fermi level approaches the band edge, the dispersion flattens, which narrows the energy window available for the rf transitions.

\begin{figure}
	\centerline{\includegraphics[clip,width=3.4in] {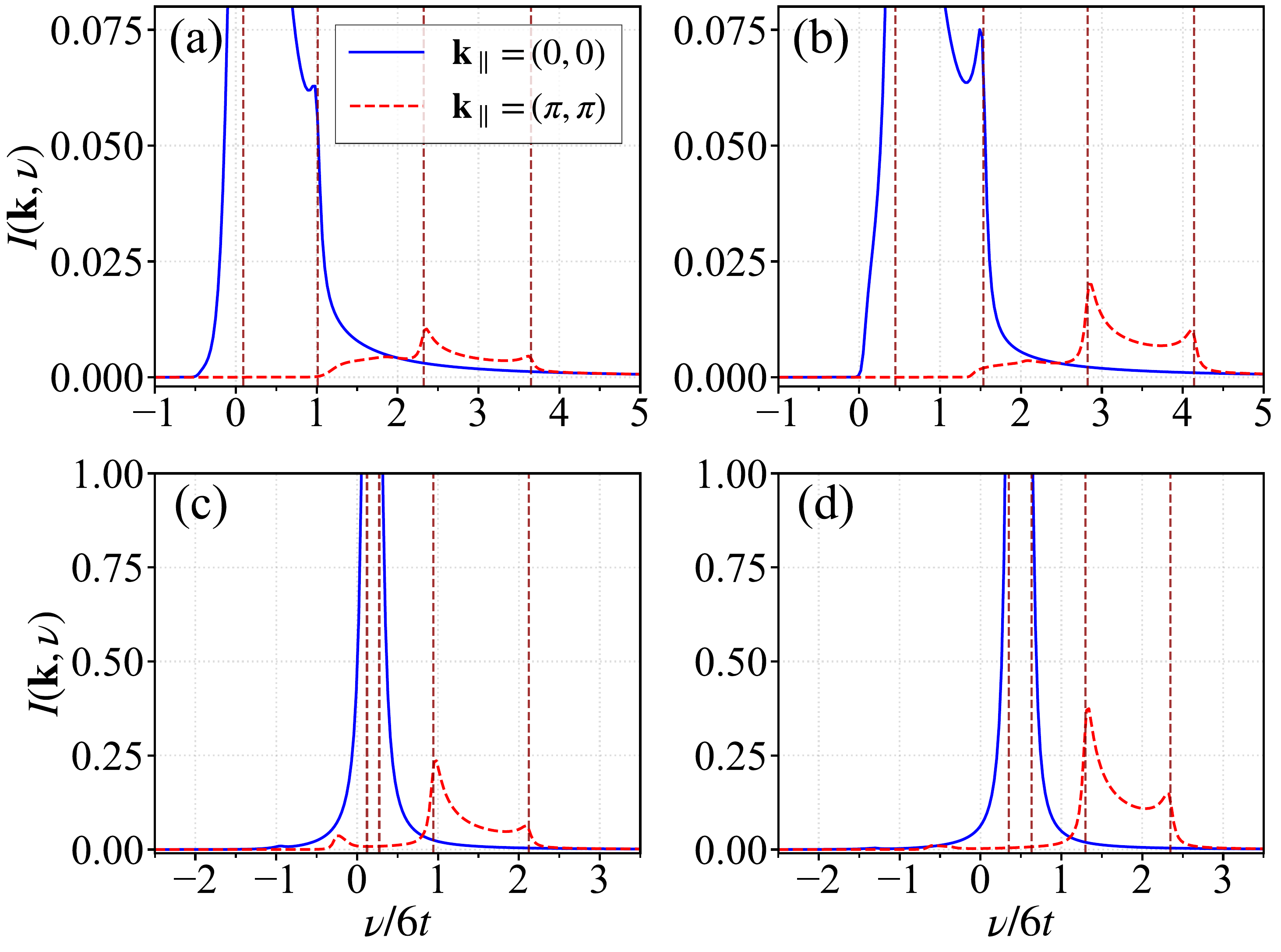}}
	\caption{EDCs $I(\mathbf{k}_{\parallel},\nu)$ at the center and edge of the first Brillouin zone for (a) $n=0.1$, $U=U_{\rm c}$, (b) $n=0.1$, $U/6t=-2$, (c) $n=1$, $U=U_{\rm c}$, and (d) $n=1$, $U/6t=-2$. The blue solid and red dashed lines denote the EDCs at $\mathbf{k}_{\parallel}=(0,0)$ and $\mathbf{k}_{\parallel}=(\pi,\pi)$, respectively. The vertical lines indicate the rf detunings corresponding to the $\mathrm{M}_{0}$--$\mathrm{M}_{3}$ vHSs. }
	\label{fig:z-integrated_MRRFEDC}
\end{figure}

To extract the vHS signatures, we plot the energy distribution curves $I(\mathbf{k}_{\parallel},\nu)$ at the zone center $\mathbf{k}_{\parallel}=(0,0)$ and zone corner $\mathbf{k}_{\parallel}=(\pi,\pi)$, as shown in Fig.~\ref{fig:z-integrated_MRRFEDC}. The EDCs at $\mathbf{k}_{\parallel}=(0,0)$ exhibit stronger overall intensity due to larger occupied spectral weight. By contrast, the EDCs at $\mathbf{k}_{\parallel}=(\pi,\pi)$ show more pronounced vHS features, which become increasingly prominent as $|U|$ and $n$ increase.

\begin{figure}
	\centerline{\includegraphics[clip,width=3.4in] {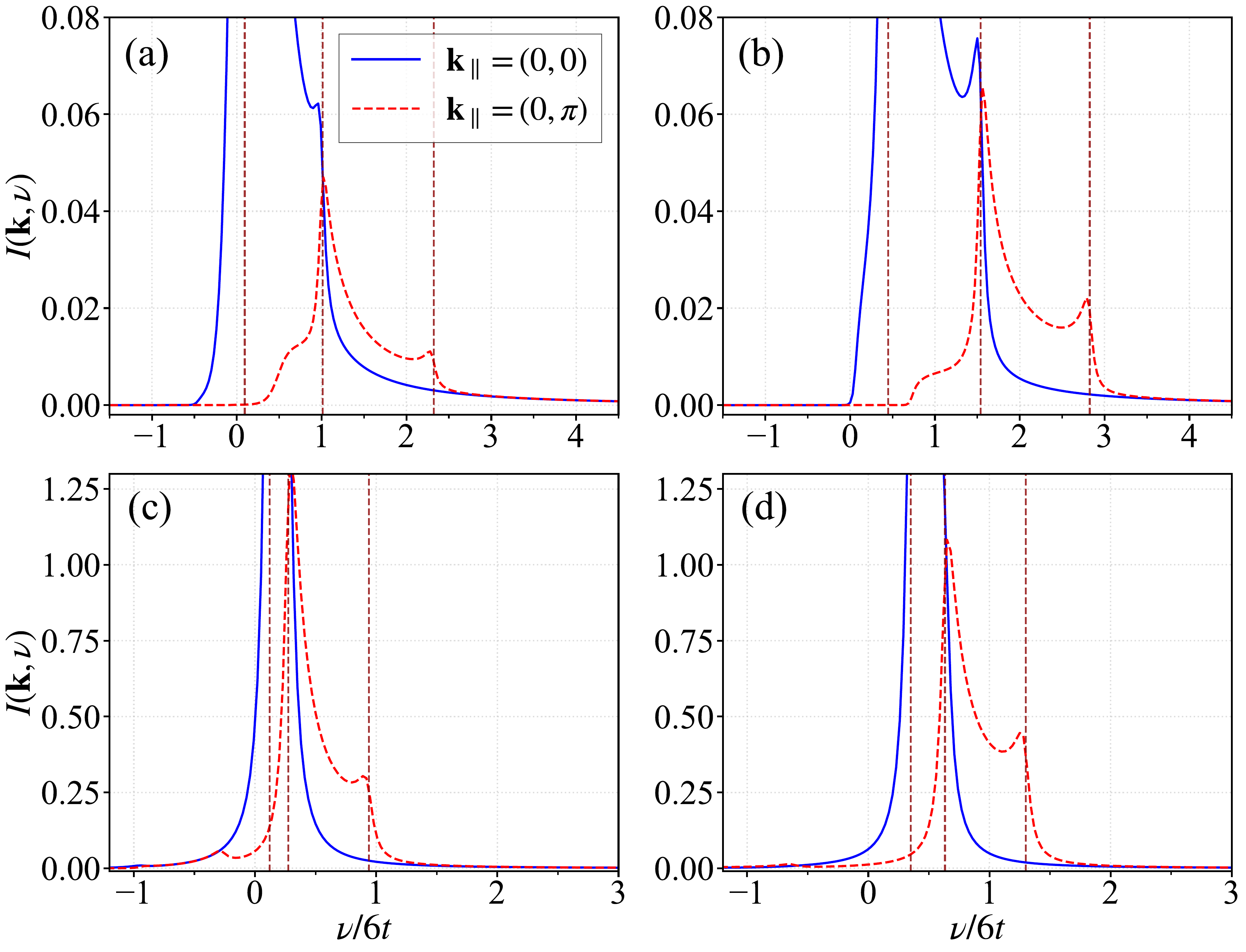}}
	\caption{EDCs $I(\mathbf{k}_{\parallel},\nu)$ with the same $(n,U)$ as in Fig.~\ref{fig:z-integrated_MRRFEDC} but at $\mathbf{k}_{\parallel}=(0,0)$ and $\mathbf{k}_{\parallel}=(0,\pi)$ in the $xy$ plane of the first BZ. The blue solid and red dashed lines denote the cases of $\mathbf{k}_{\parallel}=(0,0)$ and $\mathbf{k}_{\parallel}=(0,\pi)$, respectively. The vertical lines indicate the rf detunings corresponding to the $\mathrm{M}_{0}$--$\mathrm{M}_{2}$ vHSs. }
	\label{fig:z-integrated_MRRFEDC_X}
\end{figure}

However, the EDC signal at $(\pi,\pi)$ is weak, since $\mathrm{M}_2$ and $\mathrm{M}_3$ in Fig.~\ref{fig:AkwGaussconv} lie far above the Fermi level, so that the corresponding spectral weight, indicated by the opacity of the dispersion, is small and vanishes in the $U\rightarrow 0$ limit. To obtain stronger signals, we examine the EDCs at $\mathbf{k}_{\parallel}=(0,\pi)$, with $k_z$ integrated, for the same parameters, as shown in Fig.~\ref{fig:z-integrated_MRRFEDC_X}. For both low and high fillings, the rf signals at $\mathbf{k}_{\parallel}=(0,\pi)$ are much stronger than at $(\pi,\pi)$, since the $\mathrm{M}_1$ and $\mathrm{M}_2$ features here lie closer to or below the Fermi level. The $\mathrm{M}_1$ and $\mathrm{M}_2$ features at $\mathbf{k}_{\parallel}=(0,\pi)$ are therefore considerably more robust than the $\mathrm{M}_2$ and $\mathrm{M}_3$ features at $(\pi,\pi)$, for a given lattice size and total atom number in experiment.

\subsection{vHS detunings as functions of interaction strength}

\begin{figure}[h]
	\centerline{\includegraphics[clip,width=3.3in] {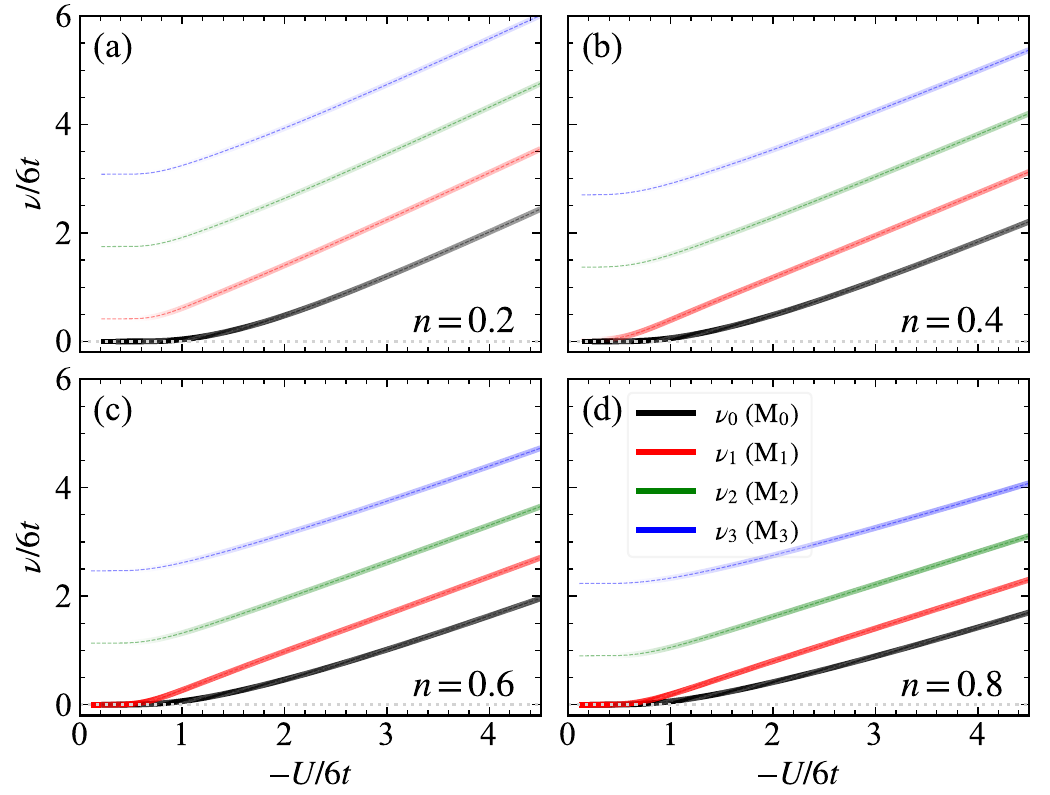}}
	\caption{vHS rf detunings $\nu_i$ ($i=0,\dots,3$) as functions of $-U/6t$ for $n=$ (a) 0.2, (b) 0.4, (c) 0.6 and (d) 0.8, computed from the kinematic relation of Eq.~(\ref{eq:fourvHS}) with the self-consistent $(\mu,\Delta)$. The opacity of the curves is coded by the spectral weight $v_\mathbf{k}^2$ at the vHS momenta.}
	\label{fig:vHS_vs_g}
\end{figure}

Figure~\ref{fig:vHS_vs_g} shows the predicted vHS rf detunings $\nu_i$ as functions of $-U/6t$, computed from Eqs.~(\ref{eq:fourvHS}) using the self-consistent $(\mu,\Delta)$ for various $n$. This plot serves as the theoretical ``map'' connecting the measurable vHS positions to the underlying pairing parameters. 

The detuning $\nu_0$ starts from zero at $U=0$, passes through an exponentially suppressed flat region, scaling approximately as $\Delta^2$, and then rises away from the horizontal axis at stronger coupling. Meanwhile, $\nu_1$, $\nu_2$, and $\nu_3$ are well separated from one another and grow monotonically with $|U|$, reflecting the increasing pairing gap, along with decreasing $\mu$. Comparing the four panels, the vHS marker positions all decrease with increasing $n$, dominated by the chemical-potential term $\epsilon_i-\mu$ in Eqs.~(\ref{eq:fourvHS}) which shifts the band energies downward. The spectral weight, in contrast, is controlled by the coherence factor $v_{\bf k}^2$ and spreads toward larger momenta as the gap grows. The separation between $\nu_1$, $\nu_2$, and $\nu_3$ across a wide coupling range indicates that multiple candidate vHS markers exist at the kinematic level. 

It should be noted, however, that the spectral weight $v_{\bf k}^2$ contributing to the $\nu_1$, $\nu_2$, and $\nu_3$ features becomes very small and vanishes at $U=0$ for low fillings, consistent with the fact that the rf spectrum in the noninteracting limit reduces to a single $\delta$-function peak at $\nu=0$.

\subsection{Extraction protocol for $\Delta$ and $\mu$}

The kinematic relations in Eqs.~(\ref{eq:fourvHS}) can be inverted algebraically. To illustrate the procedure, we outline the inversion for the $\mathrm{M}_1$ and $\mathrm{M}_2$ pair, which are well separated and experimentally favorable. Using the band energies $\epsilon_i=2i/3$ ($i=1,2$) defined in Eq.~(\ref{eq:unifiedvHS}), Eqs.~(\ref{eq:fourvHS}) can be rearranged to yield
\begin{equation}
    \Delta^2 = \nu_i(\nu_i - 2\epsilon_i + 2\mu), \qquad i=1,2.
\end{equation}
Eliminating $\Delta^2$ between the two equations gives $\mu$ directly, after which $\Delta$ follows from either equation. For concreteness, using the parameters of Fig.~\ref{fig:RFcurrent}(c) ($n=1$, $U=U_\text{c}$), the $\mathrm{M}_1$ and $\mathrm{M}_2$ detunings extracted from the spectrum yield $\mu\approx 1.0$ and $\Delta\approx 0.5$, in quantitative agreement with the self-consistent input values. Equivalent inversions can be performed for any pair of vHS features that are sufficiently resolved.

At half filling,  $\mu=1$, so that a single vHS measurement suffices:
\begin{equation}
    \Delta = \sqrt{\nu_i(\nu_i - 2\epsilon_i + 2)}.
\end{equation}
Thus one may choose the most pronounced, best separated signals among the $\rm M_1$,
$\mathrm{M}_2$ and $\mathrm{M}_3$ features for the best quantitative accuracy.

It should be noted that the pairing gap $\Delta$ represents the off-diagonal part of the self-energy, while the chemical potential $\mu$ appearing in the kinematic relations is the effective one determined by the self-consistent equations. It differs from the physical value measured relative to the noninteracting band bottom by the bare Hartree shift $nU/2$ and the other diagonal self-energy contributions. These diagonal contributions do not vanish with the gap; indeed, in the limit $\Delta\to 0$, the full Green's function $G$ reduces to the formally bare $G_0$, yet its $\mu$ may still differ from the noninteracting value.

\section{\label{sec:3}Conclusion}
We have shown that van Hove singularities in the rf spectra of the 3D attractive Fermi Hubbard model provide robust, kinematics-based markers for the pairing gap $\Delta$ and chemical potential $\mu$. Four types of vHSs were classified, and their rf detuning positions were derived as simple algebraic functions of $\Delta$ and $\mu$ alone. This makes it possible, in principle, to extract both parameters from two well-resolved and sufficiently intense singular features without full spectral dispersion fitting. The features remain identifiable under realistic lifetime and scattering broadenings, become more visible at stronger coupling where back-bending loses sensitivity, and reduce to a single-vHS diagnostic at half filling owing to particle-hole symmetry. In the $k_z$-integrated spectra, the EDCs at $\mathbf{k}_{\parallel}=(0,\pi)$ provide the strongest signals and are therefore the most favorable for experiments. We note that the present results are obtained within the pseudogap approximation $\Sigma_{\rm pg}(K)\approx -\Delta_{\rm pg}^2 G_0(-K)$, which captures the dominant $Q\approx 0$ contribution; a fully self-consistent convolution over all pair momenta would introduce additional frequency-dependent broadening, likely further softening the vHS features, and remains a challenge for future work. We also note that the present analysis employs the $G_0G$ $T$-matrix scheme; in conserving, $\Phi$-derivable ($GG$) or non-self-consistent ($G_0G_0$) formulations, the quantitative relation between vHS detunings and the pairing gap may differ, and the extent to which vHS features survive at these levels of approximation warrants further investigation. More broadly, the kinematic relations derived here hold independently of the particular many-body framework. In this sense, vHS-based gap extraction can serve as a useful cross-method benchmark, in which the parameters inferred from experiment may be compared with the values predicted by $T$-matrix theories, cluster dynamical mean-field theory, and other approaches that currently yield disparate gap estimates, thereby offering a spectroscopic means of discriminating among them. 
Recent experimental advances in 3D lattice cooling and homogeneous trapping~\cite{shao2024antiferromagnetic}, together with rf frequency resolutions reaching $0.038\,E_{\mathrm{F}}$~\cite{li2024observation,Nielsen2025}, make these predictions testable with current technology. We anticipate that vHS-based extraction will serve as a practical complement to back-bending and peak-analysis methods in ongoing quantum simulation experiments.

\section{Acknowledgments}

This work was supported by Quantum Science and Technology -- National
Science and Technology Major Project (Grant No. 2021ZD0301904).

\bibliography{References.bib}
	
\end{document}